\documentclass{article}%
\usepackage{amsmath}
\usepackage{amsfonts}
\usepackage{amssymb}
\usepackage{graphicx}
\usepackage{epstopdf}
\usepackage{cite}
\usepackage[usenames]{color}%
\providecommand{\U}[1]{\protect\rule{.1in}{.1in}}
\newcommand{\ba}{\begin{array}}
\newcommand{\ea}{\end{array}}
\newcommand{\Dsl}[1] { \setbox0=\hbox{$#1$}     
\dimen0=\wd0   \setbox1=\hbox{/} \dimen1=\wd1  \ifdim\dimen0>\dimen1        
 \rlap{\hbox to \dimen0{\hfil/\hfil}}  #1 \else \rlap{\hbox to \dimen1{\hfil$#1$\hfil}}  /  \fi  }
\newcommand{\bea}{\begin{eqnarray}}
\newcommand{\eea}{\end{eqnarray}}

\begin{document}
\noindent

\title {  \Large Production of a tensor glueball   in the reaction $\gamma\gamma\rightarrow G_2\pi^0$ at large momentum transfer }


\author{ N. Kivel \thanks{On leave of absence from St.~Petersburg Nuclear Physics
Institute, 188350, Gatchina, Russia} \  
and  M. Vanderhaeghen 
\\[3mm]
{\it Helmholtz Institut Mainz, Johannes Gutenberg-Universit\"at, D-55099
Mainz, Germany}
 \\
{\it Institut f\"ur Kernphysik, Johannes
Gutenberg-Universit\"at, D-55099 Mainz, Germany } 
 }

\date{}


\maketitle

\begin{abstract}
We study the production of a tensor glueball  in the reaction  $\gamma\gamma\rightarrow G_2\pi^0$.  We compute  the cross section at higher momentum transfer using the collinear factorisation approach.  We find that  for a value of the tensor gluon coupling of $f^T_g\sim 100$~MeV,   the cross section can  be measured in the  near future by the Belle~II experiment. 
 \end{abstract}



\section{Introduction}
\label{int}

The hadronic states   made up only  from gluons  are known as glueballs \cite{Fritzsch:1973pi}.  
The spectrum of such  states have been computed  using lattice QCD techniques, see e.g. \cite{Bali:1993fb, Morningstar:1999rf, Chen:2005mg, Richards:2010ck}. 
However,  identifying glueballs in experiment  is not an easy task because they have the same quantum numbers as quark-antiquark mesons.  At present time  there are  few scalar candidates  which can be  associated with the scalar glueballs, see e.g.  \cite{Klempt:2007cp, Crede:2008vw, Ochs:2013gi} and references therein.   Much less  is  even known about  tensor glueballs.  The lattice calculations predict the mass of the lowest  $2^{++}$ glueball state to be around  $2.3-2.4$~GeV.  There are some experimental evidences  that such state  have may been seen in various processes \cite{Etkin:1985se, Uehara:2013mbo, Ablikim:2016hlu}. 

In this Brief Report we would like to study the  production of tensor $2^{++}$ glueballs  in  two-photon  collisions  at high momentum transfer.  Despite the fact that the cross section of such process is relatively small  it   can potentially be observed at high luminosity  $e^+e^-$ colliders like Belle~II.   The production of glueballs  in the  reaction  $\gamma\gamma\rightarrow G\pi^0$  has already been studied long time ago  in Refs. \cite{Atkinson:1983yh, Wakely:1991ej,Ichola:1993yz}.  However  the tensor glueball  was considered only  in Ref.\cite{Ichola:1993yz} but within the  specific approach  where the glueball state  is described as a weakly bound state of two non-relativistic gluons.  In the present work  the production  amplitude is computed using the QCD factorisation approach \cite{Lepage:1980fj, Brodsky:1981rp, Efremov:1979qk} . The coupling of  quarks and gluons to the final mesonic states is described by the  distribution amplitudes (DAs) describing  the momentum fraction distribution of partons at zero transverse separation in a two-particle Fock state.  Our calculation is presented in Sec.\ref{calc} and results for the cross section are shown in Fig.\ref{fig-crsec}.  Concluding remarks are given in Sec.\ref{conc}.

 \section{Calculation }
 \label{calc}
The production amplitude for the $\gamma\gamma\rightarrow G\pi^0$ process can be described in terms of  the helicity  amplitudes 
 \begin{equation}
iA_{\pm\pm}=~\varepsilon_{1\mu}(\pm)\varepsilon_{2\nu}(\pm)~\int d^{4}x~e^{-i(q_{1}%
x)}\left\langle G_2(p),\pi^{0}(k)\left\vert ~T\{~J_{\text{em}}^{\mu
}(x),J_{\text{em}}^{\nu}(0)\}\right\vert 0\right\rangle 
\end{equation}
 The cross section of the process  is given by  \cite{Budnev:1974de}
\begin{equation}
\frac{d\sigma_{\gamma\gamma}[\pi^{0}G_2]}{d\cos\theta}=\frac{1}{64\pi}%
\frac{s+m^{2}}{s^{2}}\left(  |\overline{A_{++}}|^{2}+|\overline{A_{+-}}%
|^{2}\right),
\label{crsec}
\end{equation}
where  we used  $|A_{++}|=|A_{--}|$ and $|A_{-+}|=|A_{+-}|$,  and where $m$ denotes the glueball mass.  
 The square of the amplitudes in (\ref{crsec}) implies the  sum over polarisations $\lambda$  of the glueball:
\begin{equation}
|\overline{A_{+\pm}}|^{2}=\sum_{\lambda=-2}^2A_{+\pm}(\lambda)A_{+\pm} ^{\ast}(\lambda),
\label{suml}
\end{equation} 

\begin{figure}[ptb]%
\centering
\includegraphics[
height=1.4761in,
width=2.2424in
]%
{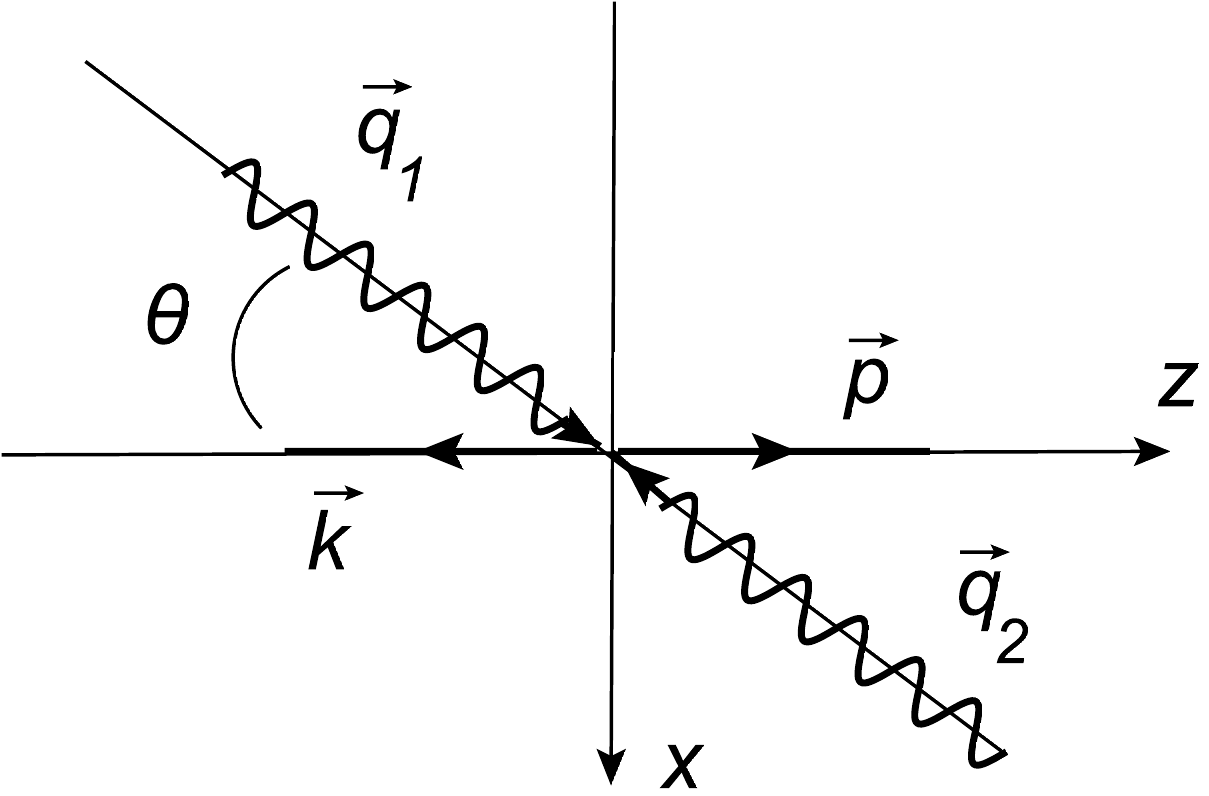}%
\caption{Kinematics of  the process $ \gamma(q_{1})\gamma(q_{2})\rightarrow
\pi(k)f_{2}(p)$.}%
\label{kinematics}%
\end{figure}

We choose the  momenta  as $ \gamma(q_{1})\gamma(q_{2})\rightarrow
\pi(k)G_{2}(p)$ and consider the  center mass system (cms)
$\boldsymbol{\bar{k}}+\boldsymbol{\bar{p}}=0$ with the pion and glueball momenta directed along
$z$-axis, see Fig.\ref{kinematics}. Let us also introduce light-like vectors $n=(1,0,0,-1)$ and $\bar n=(1,0,0,1)$  so that  
$(Vn)\equiv V_+=V_0+V_3$  and  $(V\bar n)\equiv V_-=V_0-V_3$. 
 The light-cone expansion of the particle momenta in the region with $s\sim -t\sim -u\gg \Lambda_{QCD}$ reads 
\begin{equation}
p\simeq \sqrt{s}\frac{\bar{n}}{2},~~\ \ k\simeq \sqrt{s} \frac{n}{2},
\end{equation}%
\begin{equation}
q_{1}=\sqrt{s}\frac12(1-\eta)\frac{n}{2}+\sqrt{s}\frac12(1+\eta)\frac{\bar{n}}{2}+q_{\bot},~~
q_{2}=\sqrt{s}\frac12(1+\eta)\frac{n}{2}+\sqrt{s}\frac12(1-\eta)\frac{\bar{n}}{2}-q_{\bot},\ \
\end{equation}%
\begin{equation}
q_{\bot}^{2}=\sqrt{s}\frac14(1-\eta^{2}),~\ \ \eta\equiv\cos\theta,
\end{equation}
where $\theta$ is the scattering angle in the cms, see Fig.\ref{kinematics},  and where we neglected the small  power suppressed terms. 

\begin{figure}[ptb]%
\centering
\includegraphics[width=4in]%
{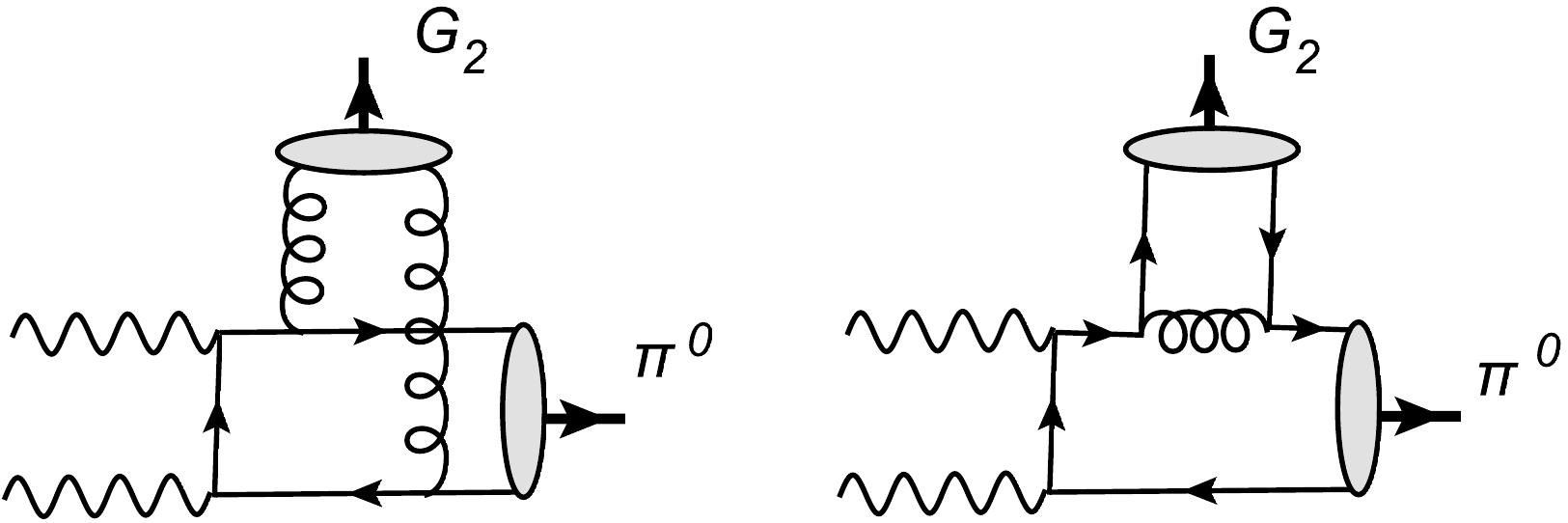}%
\caption{ Typical diagrams which describe  the amplitudes $A_{+\pm}$. The blobs denote the distribution amplitudes.  }%
\label{fig-diagrams}%
\end{figure}

In the region  $s\sim -t\sim -u\gg \Lambda_{QCD}$  the  amplitudes  can be computed in terms of convolution integrals of the hard coefficient function with the mesonic  distribution amplitudes. The typical diagrams are shown in Fig.\ref{fig-diagrams}.  The blobs in Fig.\ref{fig-diagrams}  denote  the  light-cone matrix elements which define the  DAs of the outgoing mesons.  The pion DA is defined as 
\begin{equation}
\sqrt{2}f_{\pi}\phi_{\pi}(y)=i\int\frac{d\lambda}{\pi}e^{-i\lambda
(2y-1)k_-}\left\langle \pi^{0}(k)\left\vert 
\bar{q}(\lambda\bar{n})\gamma_-\gamma_{5}q(-\lambda\bar{n})\right\vert 0\right\rangle , \label{pionDA}%
\end{equation}
where we assume for the flavor structure  $\bar{q}q=\bar{u}u-\bar{d}d$.  The pion DA has normalisation  $\int_{0}^{1}dy~\phi_{\pi}(y)=1$
 so that the pion decay constant $f_\pi$  is defined as
\begin{equation}
\left\langle \pi^{0}(k)\left\vert \overline{q}(0)~\gamma_-\gamma_{5}%
~q(0)\right\vert 0\right\rangle =-i\sqrt{2}f_{\pi}k_-,\, ~f_{\pi}=131\text{MeV}.
\end{equation}

The light-cone matrix element of tensor glueball  can be defined in the similar way as for the tensor meson $2^{++}$, see, e.g.  Ref. \cite{Braun:2016tsk}.  In general case  there are three light-cone matrix elements which define two gluon DAs and one quark DA.  In the quark  case  the distribution amplitudes is defined as   
 \begin{equation}
\langle G_{2}(p,\lambda)|\bar{q}(\lambda n)\gamma_+ q(-\lambda n)|0\rangle
=\sqrt{2}f_{q}m^{2}\frac{e_{++}^{(\lambda)}}{p_{+}}\int\limits_{0}%
^{1}\!dx\,e^{i\lambda(2x-1)p_{+}}\phi_{2}(x), \nonumber\label{phi}%
\end{equation}
where we assume for the quark flavors $\bar{q}q=\bar{u}u+\bar{d}d$.   The polarisation tensor $e_{\alpha\beta}^{(\lambda)}$ is symmetric and
traceless, and satisfies the condition $e_{\alpha\beta}^{(\lambda)}p^{\beta
}=0$\footnote{Here  $p$ is  the exact momentum  with $p^2=m^2$. }.   The polarisation sum  is given by 
\begin{equation}
\sum_{\lambda}e_{\mu\nu}^{(\lambda)} e_{\rho\sigma}^{(\lambda)\ast}=\frac{1}{2}M_{\mu\rho}M_{\nu\sigma}+\frac{1}{2}M_{\mu\sigma}M_{\nu
\rho}-\frac{1}{3}M_{\mu\nu}M_{\rho\sigma}\,, \label{polarization}%
\end{equation}
where $M_{\mu\nu}=g_{\mu\nu}-p_{\mu}p_{\nu}/m^{2}$ and the normalization condition reads $e_{\mu\nu}^{(\lambda)}e_{\mu\nu}^{(\lambda^{\prime})\ast}%
=\delta_{\lambda\lambda^{\prime}}$.
The distribution amplitude  is antisymmetric function  $\phi_{2}(1-x)=-\phi_{2}(x)$ 
and describes  transition into  the tensor glueball  with helicity $\lambda=0$.
The normalization of the  quark DA is given by 
\begin{equation}
\int\limits_{0}^{1}dx\,(2x-1)\,\phi_{2}(x)=1\,\,. 
\label{norm}%
\end{equation}
Therefore the constant $f_{q}$ is defined as a matrix element of the local operator
\begin{equation}
\langle G_{2}(p,\lambda)|\bar{q}\left\{  
\gamma_{\mu} 
(i\overset{\rightarrow}{D}_{\nu}-i\overset{\leftarrow}{D}_{\nu})
+(\mu\leftrightarrow\nu) \right\}
q|0\rangle=\sqrt{2}f_{q}m^{2}e_{\mu\nu}^{(\lambda)*}, \label{norm1}%
\end{equation}
where $D_\mu$ is the covariant derivative. 

The  gluon  DAs  are  defined as 
\begin{align}
\langle G_{2}(p,\lambda)|G_{+\mu_\bot}^{a}(\lambda n)G_{+\nu_\bot}^{a}(-\lambda
n)|0\rangle   = \int\limits_{0}^{1}\!dx\,e^{i\lambda(2x-1)p_+}
\left ( f_{g}^{T} e_{(\mu_\perp \nu_\perp )}^{(\lambda)} p_+^2 \phi_{g}^{T}(x)
 -f_{g}^{S}m^{2}\frac{1}{2}g^\bot_{\mu\nu}e_{++}^{(\lambda)} \phi_{g}^{S}(x)\right),
\end{align}
where $g^\bot_{\mu\nu}=g_{\mu\nu}-(n^\mu\bar n^\nu+n^\nu\bar n^\mu)/2$ and the short notation $e_{(\mu_\bot\nu_\bot)}^{(\lambda)}$  denotes  the transverse traceless projection 
\bea
e_{(\mu_\bot\nu_\bot)}^{(\lambda)}=  e_{\mu_\bot\nu_\bot}^{(\lambda)} -\frac
{1}{2}g^\bot_{\mu\nu}m^{2}\frac{e_{++}^{(\lambda)}}{p_+^{2}} ,\, \, \, \, g^\perp_{\mu\nu}e_{(\mu_\bot\nu_\bot)}^{(\lambda)}=0. 
 \eea
The distribution amplitudes $\phi_{g}^{T}(x)$ and $\phi_{g}^{S}(x)$ are 
symmetric with respect to the interchange of $x\leftrightarrow1-x$ and describe the
momentum fraction distribution of the two gluons having
the same and the opposite helicity, respectively.

The constants $f_{g}^{T}$ and $f_{g}^{S}$ are defined through the matrix
element of the local two-gluon operator:
\begin{align}
\langle G_{2}(p,\lambda)|G_{\alpha\beta}^{a}(0)G_{\mu\nu}^{a}(0)|0\rangle &
=f_{g}^{T}\left\{  \left[  (p_{\alpha}p_{\mu}-\frac{1}{2}m^{2}g_{\alpha\mu
})\,e_{\beta\nu}^{(\lambda)}-(\alpha\leftrightarrow\beta)\right]
-(\mu\leftrightarrow\nu)\right\} \nonumber\label{GTnorm}\\
&  +\frac{1}{2}f_{g}^{S}m^{2}\,\left\{  \left[  g_{\alpha\mu}e_{\beta\nu
}^{(\lambda)}-(\alpha\leftrightarrow\beta)\right]  -(\mu\leftrightarrow
\nu)\right\}  .
\end{align}
Using these definitions one can compute all necessary  diagrams some of which are shown in Fig.\ref{fig-diagrams}. 

The glueball with  the tensor polarisation $\lambda=\pm 2$ is only produced  if the colliding photons have the same helicities\footnote{ Note that use $\pm$ notations both for the helicity amplitudes (in case of amplitudes $A_{+\pm}$) and for the light-cone projections.  This should not lead to a  notational confusion. }   
\bea
A_{++}(s,\eta)=e_{\bot \alpha\beta}^{(\lambda)} \frac{q_{\bot}^{\alpha} 
\left (  q_{\bot}^{\beta}-\frac12 i\varepsilon^{\beta\sigma\mu\nu}q_{\bot\sigma}  n_\mu \bar{n}_\nu  \right)}
{s(1-\eta^{2})} 
\frac{f_{\pi}~f_{g}^{T}}{s}64\pi^{2}\alpha\alpha_{s}(\mu^2)(e_{u}%
^{2}-e_{d}^{2})\frac{\sqrt{2}}{N_{c}}~I_{g}^{++}(\eta),
\label{App}
\eea
where $\alpha$ is the electromagnetic coupling, $N_c$ denotes the number of colors.  The convolution integral  $I_{g}^{++}(\eta)$ reads
\begin{equation}
I_{g}^{++}(\eta)=\int_{0}^{1}dy~\frac{\phi_{\pi}(y)}{y\bar y}\int\limits_{0}^{1}\!dx\frac
{\phi_{g}^{T}(x)}{x\bar{x}}~\frac{2}{2xy-x-y+\eta(x-y)},
\label{Igpp}
\end{equation}
where we assume that $\bar x\equiv 1-x$.
The factor $(1-\eta^2)^{-1}$ which is explicitly shown  in Eq.(\ref{App}) cancels after  summation over polarisation  $\lambda$ in the  compution $|\overline{A_{++}}|^2$ defined in (\ref{suml}). Therefore the $\eta$-dependence of the cross section (\ref{crsec}) is  completely defined  by  the convolution integral $I_{g}^{++}(\eta)$.  

The production of a glueball with $\lambda=0$ (scalar polarisation)  is described   by the  amplitude $A_{+-}$ which reads 
\begin{equation}
 A_{+-}(s,\eta)=-e_{++}^{(\lambda)}\frac{m^{2}}{s}\, \frac{\,f_{\pi}}{s}~8\pi^{2}\alpha\alpha_{s}(\mu^2)(e_{u}%
^{2}-e_{d}^{2})\frac{1}{N_{c}}    \left(  C_{F} f_{q} I_{q}^{+-}
+\sqrt{2}  f_{g}^{S}  I_{g}^{+-}\right)  ,
\label{resApm}
\end{equation}%
where 
\begin{align}
~I_{q}^{+-}(\eta)  & =\int_{0}^{1}dy~\frac{\phi_{\pi}(y)}{y\bar{y}}%
\int\limits_{0}^{1}\!dx~\frac{\,\phi_{2}(x)}{x\bar{x}}\frac{\eta(1-\eta
^{2})(y-x)(1-x-y)^{2}}{ (1-x-y)^{2}(1-\eta^{2})+4x\bar{x}y\bar
{y}  }\, ,~\label{Iqpm}\\
I_{g}^{+-}(\eta)  & =\int_{0}^{1}dy\frac{\phi_{\pi}(y)}{y\bar{y}}%
\int\limits_{0}^{1}\!dx\frac{\phi_{g}^{S}(x)}{x\bar{x}}~\frac{1-2x-\eta}%
{x\bar{\eta}+y(1-2x+\eta)}\, .
\label{Igpm}
\end{align}
In order to write the convolution integrals in this form we used the symmetry properties of the DAs with respect to interchange $x\to 1-x$.  The  hard coefficient functions  for various  processes with gluons  have also been computed  in Ref.\cite{Baier:1985wv}.  We have checked up to a general factor that our results shown in Eqs.(\ref{App}) and (\ref{resApm})   are in agreement with the corresponding hard kernels in Ref.\cite{Baier:1985wv}. 
  
In order to make numerical estimates we need to specify models for the DAs and provide numerical values for the low energy glueball couplings. 
In the following we suppose that the states $f_2(2300)$ and $f_2(2340)$ which have been recently observed by the Belle \cite{Uehara:2013mbo} and BESIII \cite{Ablikim:2016hlu} collaborations are   good candidates to be tensor glueball.  
In our numerical  estimates  we use the following models of DAs. For pion  we take  
\begin{equation}
\phi_{\pi}(y)\simeq6y\bar{y}+6a_{2}(\mu)y\bar{y}C_{2}^{3/2}(2y-1),
\end{equation}
with the second moment 
\begin{equation}
a_{2}(\mu=1\text{GeV})=0.20.
\end{equation}%
This value  is close to many phenomenological estimates and  lattice QCD result \cite{Braun:2015axa}. 

For the glueball  DAs we take the simplest asymptotic  models 
\begin{equation}
\phi_{2}(x)=30x\bar{x}(2x-1),
\end{equation}%
\begin{equation}
\phi_{g}^{T}(x)=\phi_{g}^{S}(x)=30x^{2}\bar{x}^{2}.
\end{equation}

\begin{figure}[ptb]%
\centering
\includegraphics[
height=1.2528in,
width=5.7783in
]%
{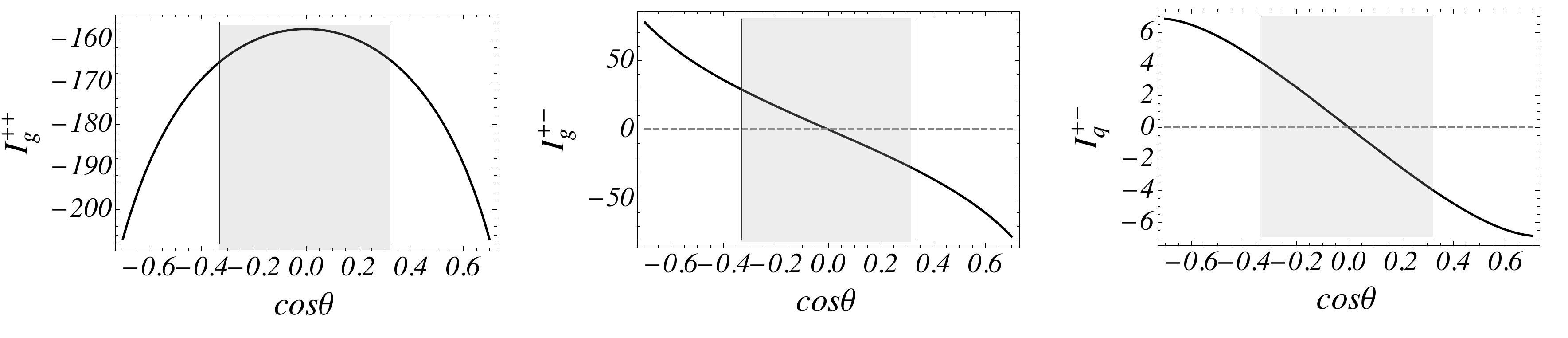}%
\caption{The convolution integrals as a functions of $\cos\theta$.    The shaded
area between the vertical lines corresponds to the region where $|u|,|t|\geq
2.5$~GeV$^{2}$ for $s=13$~GeV$^{2}$. The factorisation scale is fixed to be  $\mu^{2}=3.2$~GeV$^{2}$. For the model of the DAs used to make these predictions, see text. }%
\label{fig-iconv}%
\end{figure}

Let us first  consider the properties of the convolution integrals $I_{g,q}^{+\pm}$ (of Eqs. (\ref{Igpp}), (\ref{Iqpm}) and (\ref{Igpm})).  In
Fig.\ref{fig-iconv} we show the values of the convolution integrals as a
function of $\cos\theta$.  Below we assume that factorisation  works  reasonably for such values of $\theta$  where $|u|,|t|\geq2.5$~GeV$^{2}$. 
This region corresponds to the inner area between the  two vertical
lines on the plots in Fig.\ref{fig-iconv}.  One can easily see that in the vicinity of $\theta =90^o$ 
\begin{equation}
|I_{g}^{++}|\gg|I_{g}^{+-}|\gg|I_{q}^{+-}|.\label{convI}%
\end{equation}%
The integrals  $I^{+-}_{g,q}$ vanish  at  $\theta=90^{o}$  
therefore  we can conclude that  at least around this  point  the dominant contribution will be given by  the  amplitude $A_{++}$  which describes the production of the glueball  in the tensor polarisation. 

The values of the couplings $f^{T,S}_g$ and $f_q$  are  not  known.  It is natural  to assume that the glueball state   strongly  overlaps  with the gluon wave function and the value of the gluon couplings  are  relatively large and can be  of  the same order as the quark coupling $f_q\sim 100$~MeV  for quark-antiquark mesons, i.e.   $ f^{S}_{g}\sim f_{g}^{T}\sim 100$~MeV.   For the glueball quark  coupling $f_q$  we consider  the different scenarios  with $f_q\ll f_{g}$ and $f_q\sim  f_{g}$   corresponding to the small and to the large quark-antiquark component, respectively.  Such scenarios  will be described  by  the following numerical values
\begin{equation}
f_{q}(\mu=1~\mathrm{GeV})\simeq10-100~\mathrm{MeV}\,,
\label{fqnum}
\end{equation}
\begin{equation}
f_{g}(\mu=1~\mathrm{GeV})\simeq100~\mathrm{MeV},
\label{fgnum}
\end{equation}
\begin{equation}
f_{g}^{T}(\mu=1~\mathrm{GeV})\simeq50-150~\mathrm{MeV.}
\label{fTg}
\end{equation} 
The evolution  of these coupling  is the same as the evolution of the corresponding coupling for the tensor meson $f_2(1270)$ except for flavor mixing  and can be found in Ref.\cite{Braun:2016tsk}.  Let us notice that the tensor gluon DA  $\phi^T_g$ does not mix under evolution  with  quark  contributions  and therefore it describes the genuine gluon component of the glueball wave function.   
  
The numerical estimates show  that  the value of the cross section is practically saturated by the contribution from the amplitude
$A^{++}$ describing the production of a glueball in the  tensor polarisation.  The contribution  of the amplitude $|A^{+-}|$  is always about 
two orders of  magnitude smaller for  all numerical values  of the couplings
$f_{q}$ and $f_{g}$ shown in Eqs.(\ref{fqnum}) and (\ref{fgnum}). Therefore we can conclude  that  the contribution with $|A^{+-}|$  does not provide 
 significant  numerical impact.  Hence  the cross section is only sensitive to the value of tensor coupling  $f_{g}^{T}$.  This can also be seen, for instance,  from  the  analysis of the decay $G_2\rightarrow \phi\phi$ which can be used  for  identification of the glueball state.

\begin{figure}[ptb]%
\centering
\includegraphics[
height=1.8265in,
width=5.5475in
]%
{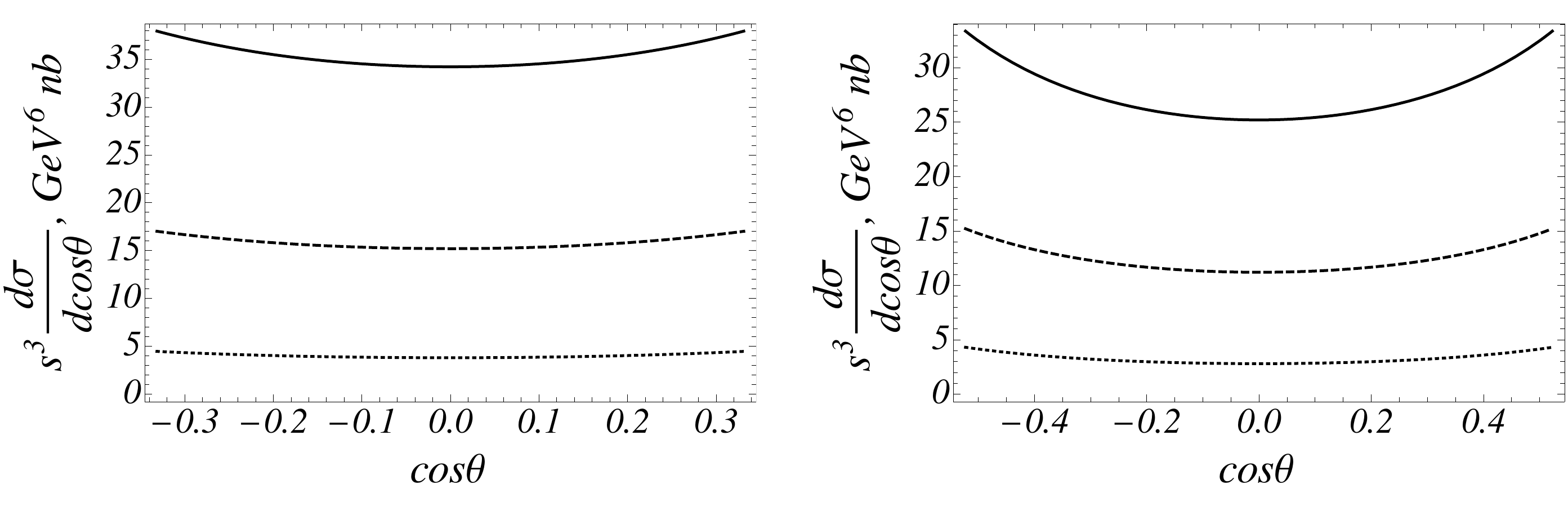}%
\caption{The cross section as a function of $\cos\theta$ at $s=13$~GeV$^{2}$
(left) and $s=16$~GeV$^{2}$ (right) in the region $|t|,|u|\geq2.5$~GeV$^{2}$.
 The solid, dashed  and dotted  lines correspond to $f_{g}^{T}(1$%
GeV$)=150,100,50$~MeV, respectively. }%
\label{fig-crsec}%
\end{figure}

In Fig.\ref{fig-crsec} we show the cross section as
a function of $\cos\theta$  at fixed values of energy $s$.    In the numerical calculations 
we take $n_{f}=3$ and  $\alpha_{s}(m_\tau^2)=0.297$.   The cross section are shown for the energy values  $s=13$~GeV$^2$ and
$16$~GeV$^{2}$.  The factorisation  scale is fixed to be   $\mu^{2}=3.2$~GeV$^{2}$ and $\mu^{2}=4$~GeV$^{2}$, respectively.  
The  values of  $\cos\theta$ correspond to the restriction  $|t|,|u|\geq2.5$~GeV$^{2}$.
 We obtain  that  for $f_{g}^{T}(1$~GeV~$)=100$~MeV 
 \begin{equation}
 s^3 \frac{d\sigma}{d \cos\theta}\simeq 11-17\, \text{GeV}^3\text{nb}.
 \end{equation}
 
\begin{figure}[ptb]%
\centering
\includegraphics[width=3.5in]%
{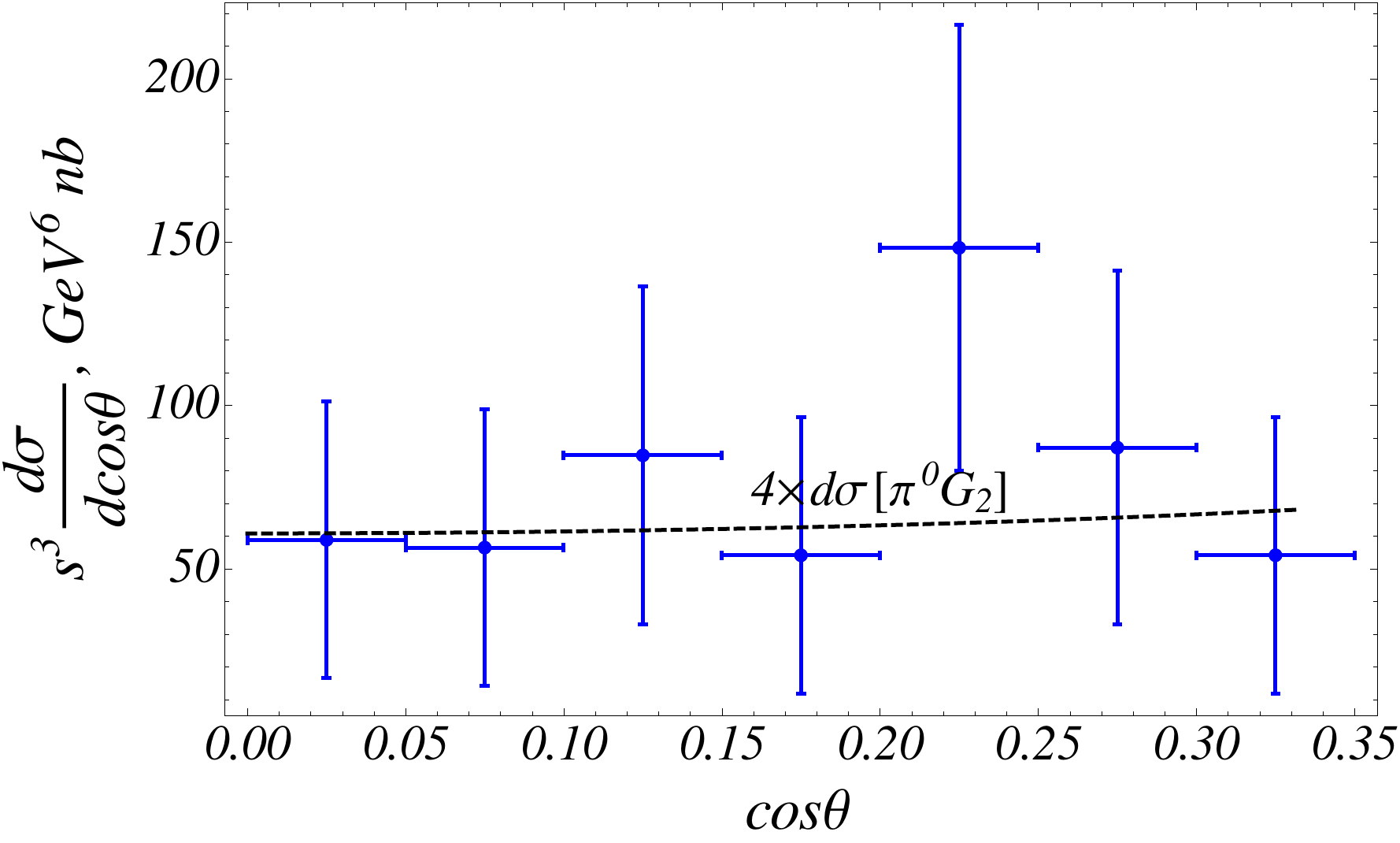}%
\caption{ Comparison of the  glueball cross section (the dashed line is the same as in Fig.\ref{fig-crsec} but scaled by factor 4 ) and data for the $\pi^0\pi^0$ cross section  for $s=13$~GeV$^2$. The data are taken from Ref.\cite{Uehara:2009cka}  }%
\label{fig-crsec-comp}%
\end{figure}

  In Fig.\ref{fig-crsec-comp} we show the glueball cross section for  $f_{g}^{T}(1$~GeV$)=100$~MeV  and  $s=13$~GeV$^2$ in  comparison  with the cross section  data for  $\gamma \gamma\rightarrow\pi^{0}\pi^{0}$  for  $s=13.3$~GeV$^2$. The data are taken from  Ref. \cite{Uehara:2009cka}.   
For  convenience the glueball cross section is scaled  by a factor 4. 
From this picture one can conclude  that the measurement  of  $\gamma
\gamma\rightarrow G_2\pi^{0}$ cross section requires a larger
luminosity which will be  achieved  in the  Belle~II experiment. 

\section{Conclusions}
\label{conc}
We calculated the amplitudes and  cross sections for the production  of a tensor glueball  in the  reaction $\gamma\gamma\rightarrow G_2\pi^0$.  We obtained that for  the  value  of the low energy coupling  $f_g^T\simeq 100$~MeV the  cross section  is dominated by the  contribution describing the production of a glueball  in the  tensor polarisation. A corresponding measurements allow one to constrain  the value of the tensor coupling $f^T_g$.  We  expect  that  the corresponding  cross section  can be  observed in the upcoming higher statistic  Belle~II experiment.  


\end{document}